\documentclass[
	a4paper, % Paper size, use either a4paper or letterpaper
	10pt, % Default font size, can also use 11pt or 12pt, although this is not recommended
	unnumberedsections, % Comment to enable section numbering
	twoside, % Two side traditional mode where headers and footers change between odd and even pages, comment this option to make them fixed
]{LTJournalArticle}

\usepackage{color, colortbl}
\definecolor{shadecolor}{gray}{0.9}
\usepackage{tabularx}
\usepackage{multirow}
\usepackage{threeparttable}
\usepackage{booktabs}
\usepackage{comment}
\usepackage[acronym]{glossaries}

\glsdisablehyper
\newacronym{SBO}{SBO}{Stack Buffer Overflow}
\newacronym{ML}{ML}{Machine Learning}
\newacronym{PULP}{PULP}{Parallel Ultra Low Power}
\newacronym{HPC}{HPC}{Hardware Performance Counter}
\newacronym{AES}{AES}{Advanced Encryption Standard}
\newacronym{RSA}{RSA}{Rivest–Shamir–Adleman}
\newacronym{SHA}{SHA}{Secure Hash Algorithm}
\newacronym{OC-SVM}{OC-SVM}{One-class Support Vector Machines}
\newacronym{LOF}{LOF}{Local Outlier Factor}
\newacronym{IF}{IF}{Isolation Forest}
\newacronym{EE}{EE}{Elliptic Envelope}

\runninghead{} % A shortened article title to appear in the running head, leave this command empty for no running head

\footertext{\textit{RISC-V Summit Europe, Munich, 24-28th June 2024}} % Text to appear in the footer, leave this command empty for no footer text

\title{Hardware-based stack buffer overflow attack detection on RISC-V architectures} % Article title, use manual lines breaks (\\) to beautify the layout

\author{%
	Cristiano Pegoraro Chenet\textsuperscript{1},
        Ziteng Zhang\textsuperscript{1},
        Alessandro Savino\textsuperscript{1},
        Stefano Di Carlo\textsuperscript{1}
        \thanks{Corresponding author: \href{mailto:stefano.dicarlo@polito.it}{\tt stefano.dicarlo@polito.it}}
}

\date{\footnotesize\textsuperscript{\textbf{1}}Politecnico di Torino}

\renewcommand{\maketitlehookd}{%
	\begin{abstract}
		\noindent This work evaluates how well hardware-based approaches detect stack buffer overflow (SBO) attacks in RISC-V systems. We conducted simulations on the PULP platform and examined micro-architecture events using semi-supervised anomaly detection techniques. The findings showed the challenge of detection performance. Thus, a potential solution combines software and hardware-based detectors concurrently, with hardware as the primary defense. The hardware-based approaches present compelling benefits that could enhance RISC-V-based architectures.
	\end{abstract}
}

\begin{document}

\maketitle % Output the title section

%----------------------------------------------------------------------------------------
%	ARTICLE CONTENTS
%----------------------------------------------------------------------------------------

\glsresetall

\section{Introduction}

Cybersecurity is paramount on a global scale. The World Economic Forum Global Risk Report 2023 ranks cyber insecurity eighth among top global risks, alongside threats like climate change and involuntary migration \cite{GlobalRisckReport2023}. Attacking software systems by exploiting memory-corruption vulnerabilities is one of today's most common attack methods \cite{Brohet_Regazzoni_2023}. Frequently, this vulnerability is exploited to redirect the program's execution flow, enabling arbitrary code execution. The \gls{SBO} attack is a notable case characterized by a nonvalidated input overflowing a buffer allocated in the memory stack. In its canonical form, the function return address in the stack is overwritten, deviating the execution to a malicious function.

Previous research proposed detecting security breaches based on hardware events \cite{Demme_2013}. The idea involves dynamically analyzing micro-architecture events in a processor using \gls{ML} algorithms. These approaches have exciting strengths: the possibility of runtime detection, adaptability to code variants and zero-day breaches, resilience against subverting the protection mechanism, and reduced detection costs \cite{Chenet_2023}.

In this work, we analyze the performance of the hardware-based approaches in detecting security breaches in RISC-V, specifically \gls{SBO} attacks. We focus on semi-supervised anomaly detection, which offers two advantages: (i) it does not require a security breach dataset for training, and (ii) it can detect zero-day breaches. Moreover, we also evaluate four different classification algorithms and an autoencoder. This last is a feedforward neural network trained to learn the most salient features of the data, thus improving the detection of traditional \gls{ML} algorithms.

%------------------------------------------------

\section{Methodology}

\begin{figure}[ht]
  \centering
  \includegraphics[width=\linewidth]{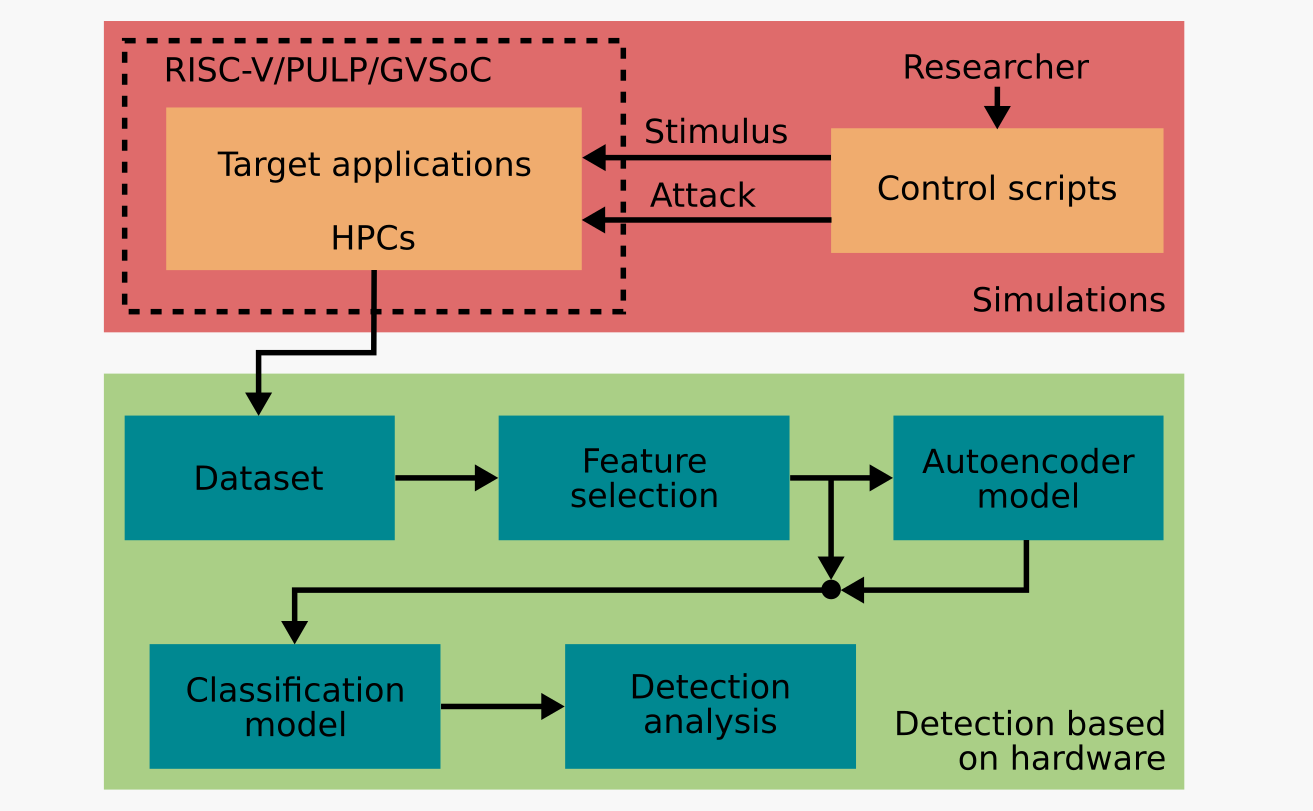}
  \caption{Methodology overview.}
  \label{fig:methodology_overview}
\end{figure}

The methodology comprises simulations and detection of security breaches based on hardware events, according to Figure \ref{fig:methodology_overview}.

Control scripts running on the host operating system automate the simulation. They properly stimulate the target applications and inject the attacks. The target applications contain benchmarks, memory-corruption vulnerabilities, and malicious code that run bare-metal in the \gls{PULP} platform through the GVSoC simulator \cite{Bruschi_2021}. \glspl{HPC} read by the target applications reflect the hardware micro-architecture events.

Simulations are performed running each target application several times, with the benchmarks stimulated with random (in dimension and value) input words. The \glspl{HPC} are just recorded at the end of the application execution. Thanks to the \gls{ML} techniques employed, the detection of breaches is allowed with any input in the benchmark once the classifiers are trained with data from a wide range of inputs. Although this choice implies a challenge in attack detection, it reduces the performance overhead imposed on the host system.

A dataset with \glspl{HPC} collected was built, and feature selection was applied. The \glspl{HPC} we leveraged from the \gls{PULP} RI5CY core reflect the number of instructions executed, load data hazards, data memory loads executed, data memory stores executed, unconditional jumps, branches (taken and not taken), taken branches and compressed instructions executed. Even if our dataset has few features, feature selection is needed for ranking purposes, considering that in real scenarios, cores may have strict limitations on the number of \glspl{HPC} recorded at a time.

Four classification models were employed through the Python Scikit-learn library: \gls{OC-SVM}, \gls{LOF}, \gls{IF}, and \gls{EE}. After performance analysis with these traditional classifiers, the autoencoder (through Keras library) was inserted in the path to evaluate the gain of this technique in the detection.

%------------------------------------------------

\section{Experimental results}

The detection of security breaches through a hardware-based approach is demonstrated for \gls{SBO} attacks. In the target applications, the benchmarks employed were \gls{AES}, \gls{RSA} encryption, \gls{SHA}, and Dijkstra algorithm. Buffer overflow vulnerabilities were artificially inserted into them, emulating potential memory corruption vulnerabilities that programs may have. After the \gls{SBO} attack succeeds, the program control flow is deviated to a Fibonacci number generator, exemplifying an anomaly. For each target application, 10k executions were performed without \gls{SBO} attacks (building the training dataset) and 10k mixing without and with attacks (building a balanced testing dataset).

\begin{figure}[ht]
  \centering
  \includegraphics[width=\linewidth]{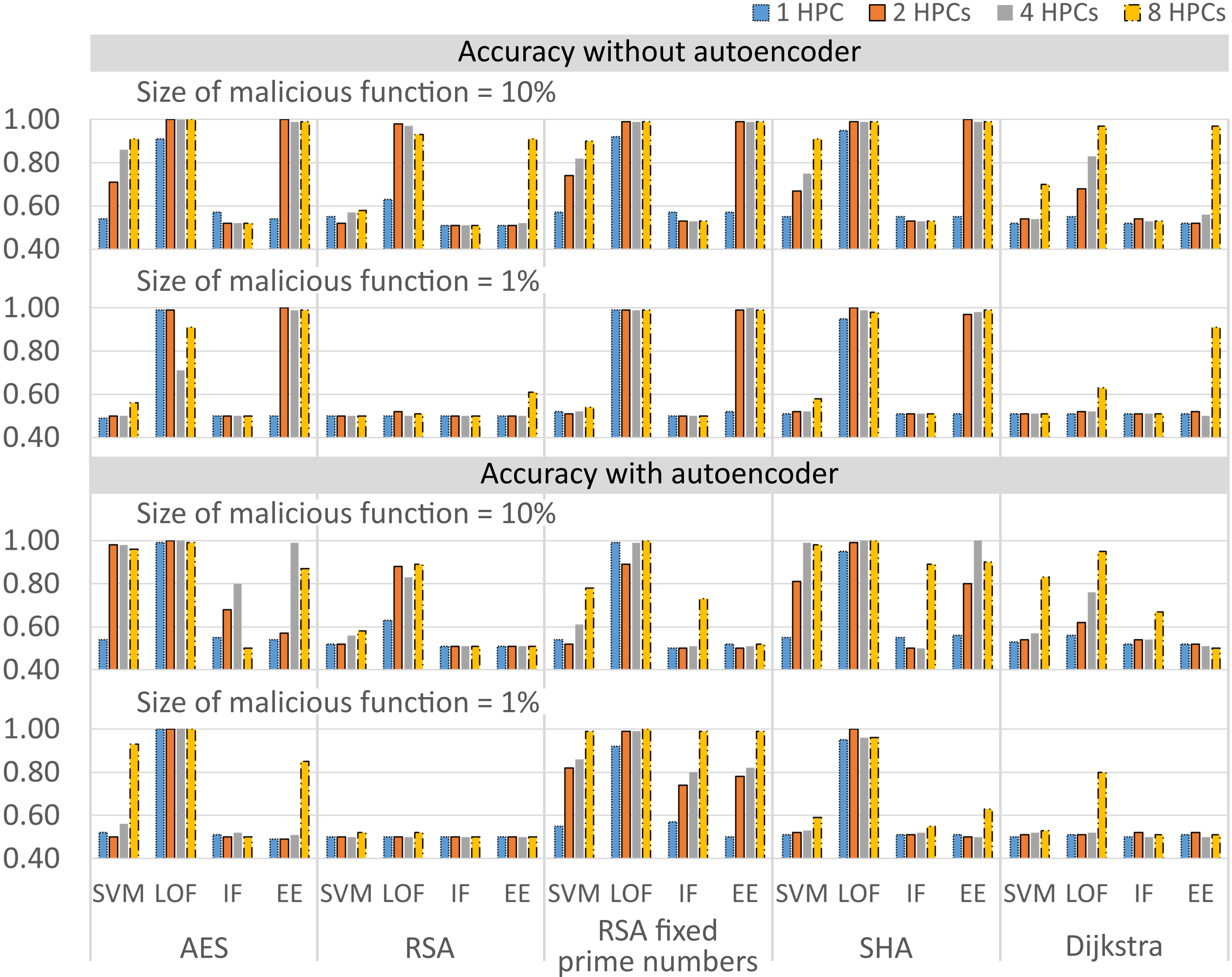}
  \caption{Performance obtained. SVM is \gls{OC-SVM}.}
  \label{fig:performance}
\end{figure}

Figure \ref{fig:performance} presents the performance obtained as a function of the malicious function size, expressed as a percentage of the number of instructions executed by the benchmark application. Without autoencoder, in \gls{AES}, \gls{RSA} (with fixed prime numbers), \gls{SHA}, and Dijkstra, a malicious function size of 1\% is enough to an accuracy higher than 90\%. The reduced detection performance with the complete \gls{RSA} stems from the random search for prime numbers when the algorithm generates the public and private keys. Moreover, it is interesting to note the good performance of the \gls{LOF} classifier: it has accuracy of at least 95\% with just 1 \gls{HPC} and malicious function size of 1\% (AES, RSA with fixed prime numbers and SHA applications). Conversely, we may note that there is no significant gain with the addition of the autoencoder.

%------------------------------------------------

\section{Final considerations}

The detection of security breaches based on hardware events is an open field, showing compelling benefits that could enhance RISC-V-based architectures, like runtime detection, adaptability to code variants and zero-day breaches, resilience, and reduced detection costs. The detection performance is the main challenge in the approach. Thus, a potential solution combines software and hardware-based detectors concurrently, with hardware as the primary defense.

%----------------------------------------------------------------------------------------
%	 REFERENCES
%----------------------------------------------------------------------------------------

\printbibliography % Output the bibliography

\scriptsize{This work was partially supported by project SERICS (PE00000014) under the MUR National Recovery and Resilience Plan funded by the European Union - NextGenerationEU and by the Vitamin-V project (Project number: 101093062) funded by the European Union. Views and opinions expressed are, however, those of the author(s) only and do not necessarily reflect those of the European Union or the HaDEA. Neither the European Union nor the granting authority can be held responsible for them.}

%----------------------------------------------------------------------------------------

\end{document}